\input{aipcheck}

\documentclass[
    ,final            
  ]
  {aipproc}

\layoutstyle{6x9}

\def\apj{ApJ}
\def\aap{A\&A}
\def\mnras{MNRAS}

\begin{document}

\title{The true redshift distribution of Pre-SWIFT gamma-ray bursts}


\author{B. Gendre}{
  address={IASF/INAF, via fosso del cavaliere 100, 00133 Roma, Italy}
}

\author{M. Bo\"er}{
  address={Observatoire de Haute Provence, 04870 Saint Michel l'Observatoire, France}
}

\begin{abstract}
SWIFT bursts appear to be more distant than previous bursts. We present the Boer \& Gendre relation that link 
redshift and afterglow luminosities. Taking advantage of the XMM-Newton, Chandra and BeppoSAX catalogs, and 
using this relation, we have investigated the redshift distribution of GRBs. We find that XMM burst sources 
with unknown redshift appear to be more distant than those with a known redshift. We propose that this effect 
may be due to a selection effect of pre-SWIFT optical observations.

\end{abstract}

\maketitle


\section{Introduction}

 The observations of long Gamma-Ray Burst (GRB) afterglows allowed the emergence of the fireball model 
\cite{ree92, mes97, pan98}. In this model an isotropic blast wave propagates into a surrounding uniform 
interstellar medium (ISM). Two refinements were made: first, the isotropic assumption was relaxed. This model 
was called the "jet model" \cite{rho97}. Second, observations showed that long GRBs may be linked with the 
explosion of a massive star (hypernova, \cite{mes01}). In such a case, the surrounding medium is not uniform 
\cite{che99} because of the wind  from the progenitor of the GRB. This model is referred as the "wind model" 
\cite{dai98, pan98, che99}.

\begin{figure}
  \includegraphics[height=.3\textheight]{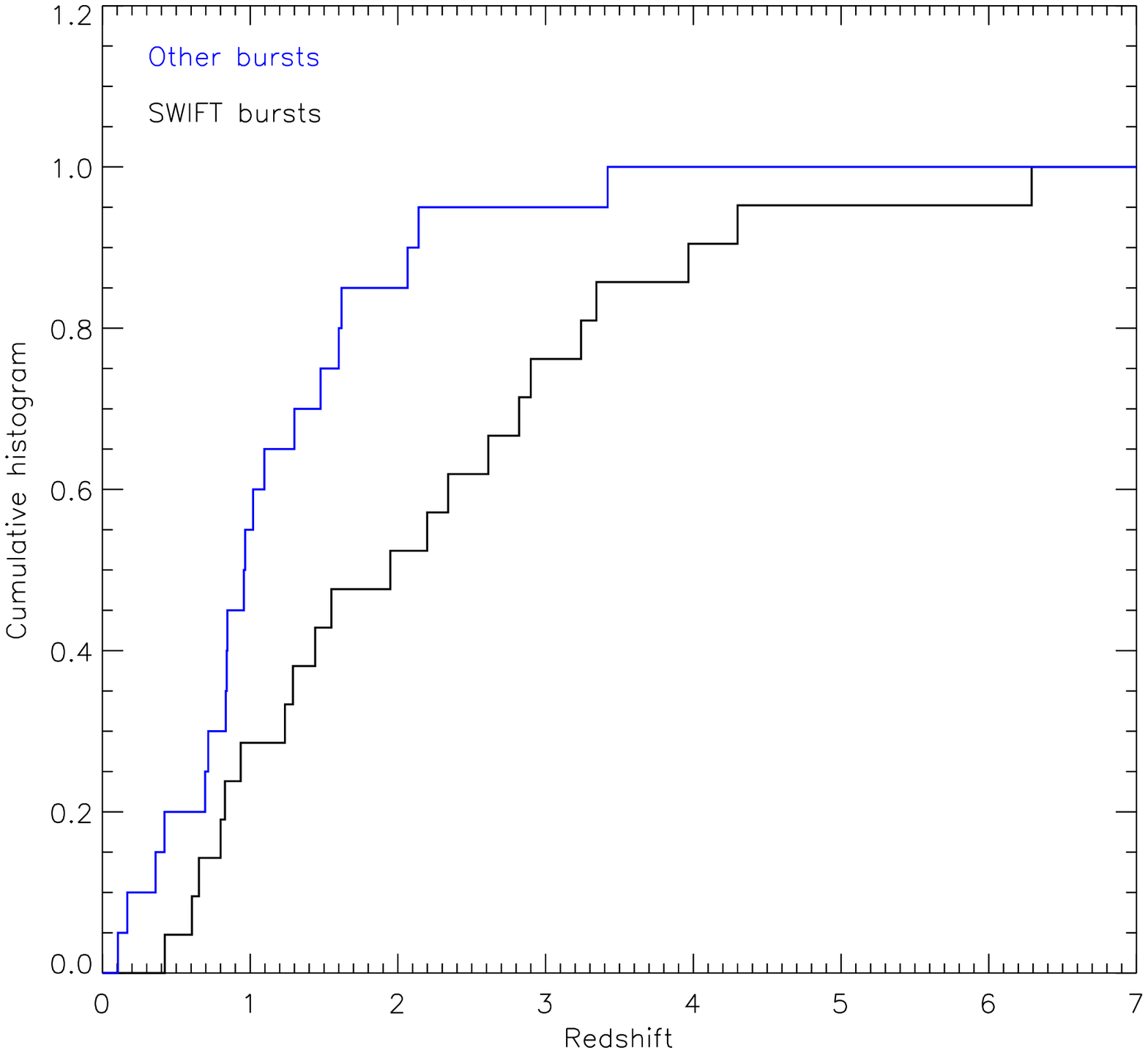}
  \includegraphics[height=.3\textheight]{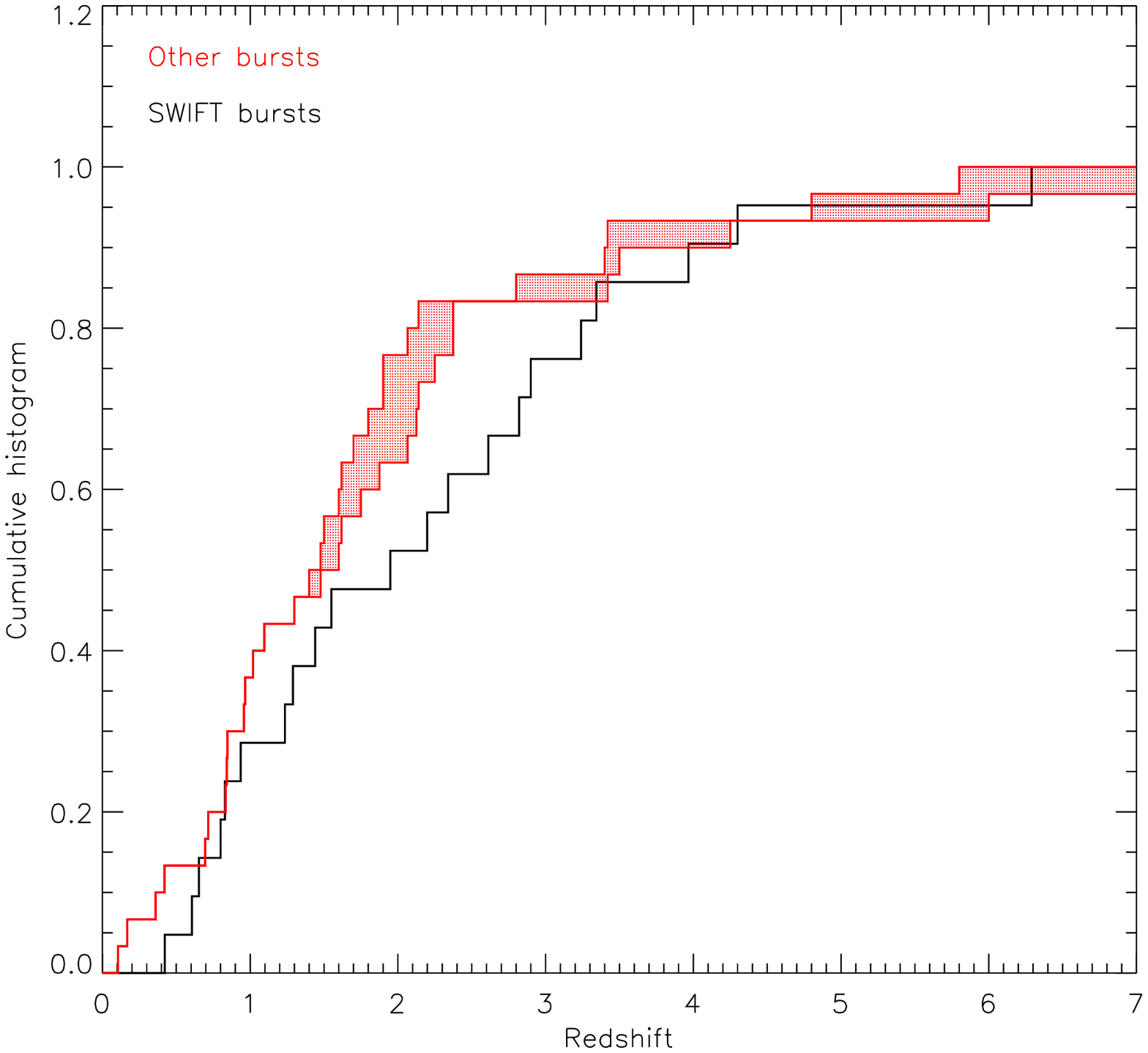}
  \caption{\label{fig0} The redshift distributions of pre-SWIFT (color line) and SWIFT (black line) bursts. 
Left : using only bursts with a redshift measured by optical studies. Right : using bursts with a redshift 
measured either by an optical study or by the use of the Bo\"er \& Gendre relation. The shaded area 
represents the uncertainties of the redshift distribution.}
\end{figure}

As GRBs are distant events, one may use them for cosmological studies. They may trace the star formation 
history and constrain the cosmological parameters. With distant bursts, such as GRB 050904 (z=6.29, [8]), 
observed by SWIFT and TAROT \cite{cus06, boe06b}, one can also observe the death of the first stars and the 
re-ionization period. Recently, it appears that SWIFT bursts were observed to be more distant than previous 
bursts \cite{jak05}. This is rather obvious when looking at Fig. \ref{fig0}, which displays the cumulative 
distribution of redshifts derived from pre-SWIFT observations versus those measured from SWIFT detections.

We reported earlier that GRB X-ray afterglows with known redshifts have a bimodal luminosity evolution : the 
faintest GRB afterglows appear to decay more slowly than the brighter ones \cite{boe00}. Bright and faint 
X-ray afterglows are separated by one order of magnitude in flux one day after the burst.We can use this 
relation
as a redshift estimator. From the distances we compute, we are able to explain the observed difference in the 
cumulative distributions of redshift presented in Fig. \ref{fig0}, as a selection bias.

\section{The Bo\"er \& Gendre relation}

\subsection{Derivation of the relation}

\begin{table}
\begin{tabular}{ccc|ccc}
\hline
 \tablehead{1}{r}{b}{Burst \\ name} & \tablehead{1}{r}{b}{Redshift}
  & \tablehead{1}{r}{b}{X-ray \\ satellite}  & \tablehead{1}{r}{b}{Burst \\ name} & 
\tablehead{1}{r}{b}{Redshift}
  & \tablehead{1}{r}{b}{X-ray \\ satellite} \\
\hline
GRB 970228 & 0.695 & BeppoSAX & GRB 011121 & 0.36  & BeppoSAX \\
GRB 970508 & 0.835 & BeppoSAX & GRB 011211 & 2.14  & XMM-newton \\
GRB 971214 & 3.42  & BeppoSAX & GRB 020405 & 0.69  & Chandra \\
GRB 980425 & 0.0085& BeppoSAX & GRB 020813 & 1.25  & Chandra \\
GRB 980613 & 1.096 & BeppoSAX & GRB 021004 & 2.3   & Chandra \\
GRB 980703 & 0.966 & BeppoSAX & GRB 030226 & 1.98  & Chandra \\
GRB 990123 & 1.60  & BeppoSAX & GRB 030328 & 1.52  & Chandra \\
GRB 990510 & 1.619 & BeppoSAX & GRB 030329 & 0.168 & XMM-Neton \\
GRB 991216 & 1.02  & Chandra  & GRB 031203 & 0.105 & XMM-Newton \\
GRB 000210 & 0.846 & BeppoSAX & GRB 050401 & 2.90  & SWIFT \\
GRB 000214 & 0.42  & BeppoSAX & GRB 050525A& 0.606 & SWIFT \\
GRB 000926 & 2.066 & BeppoSAX & GRB 050904 & 6.29  & SWIFT \\
GRB 010222 & 1.477 & BeppoSAX & GRB 050908 & 3.344 & SWIFT \\
\hline
\end{tabular}
\caption{Bursts used to derive the Bo\"er \& Gendre relation. We indicate for each burst its redshift and the 
satellite that observed the X-ray afterglow.}
\label{table1}
\end{table}

Our sample is listed in Table \ref{table1}. We used only GRBs with known redshifts that exhibit an X-ray 
afterglow observed either by BeppoSAX, XMM-Newton, Chandra or SWIFT. The detail of data analysis is presented 
in \cite{gen05}.

We have corrected the fluxes for distance, time dilation, and energy losses due to the cosmological energy 
shift. To compute these corrections, we used a flat universe model, with an $\Omega_m$ value of 0.3. We 
normalized the flux to a common distance of z=1 rather than using the luminosity. We corrected the 
cosmological energy shift as in \cite{lam00}. In order to reduce uncertainties, we did not correct for the 
time dilation effect by interpolating the flux as in \cite{lam00}; instead, we computed the time of the 
measurement in the burst rest-frame. Finally, we restricted the light curves to the 2.0$-$10.0 keV X-ray 
band, where the absorption is negligible. This allowed us to neglect any other corrections for absorption by 
the ISM. We do not take into account any beaming due to a possible jet.

\begin{figure}
  \includegraphics[height=.27\textheight]{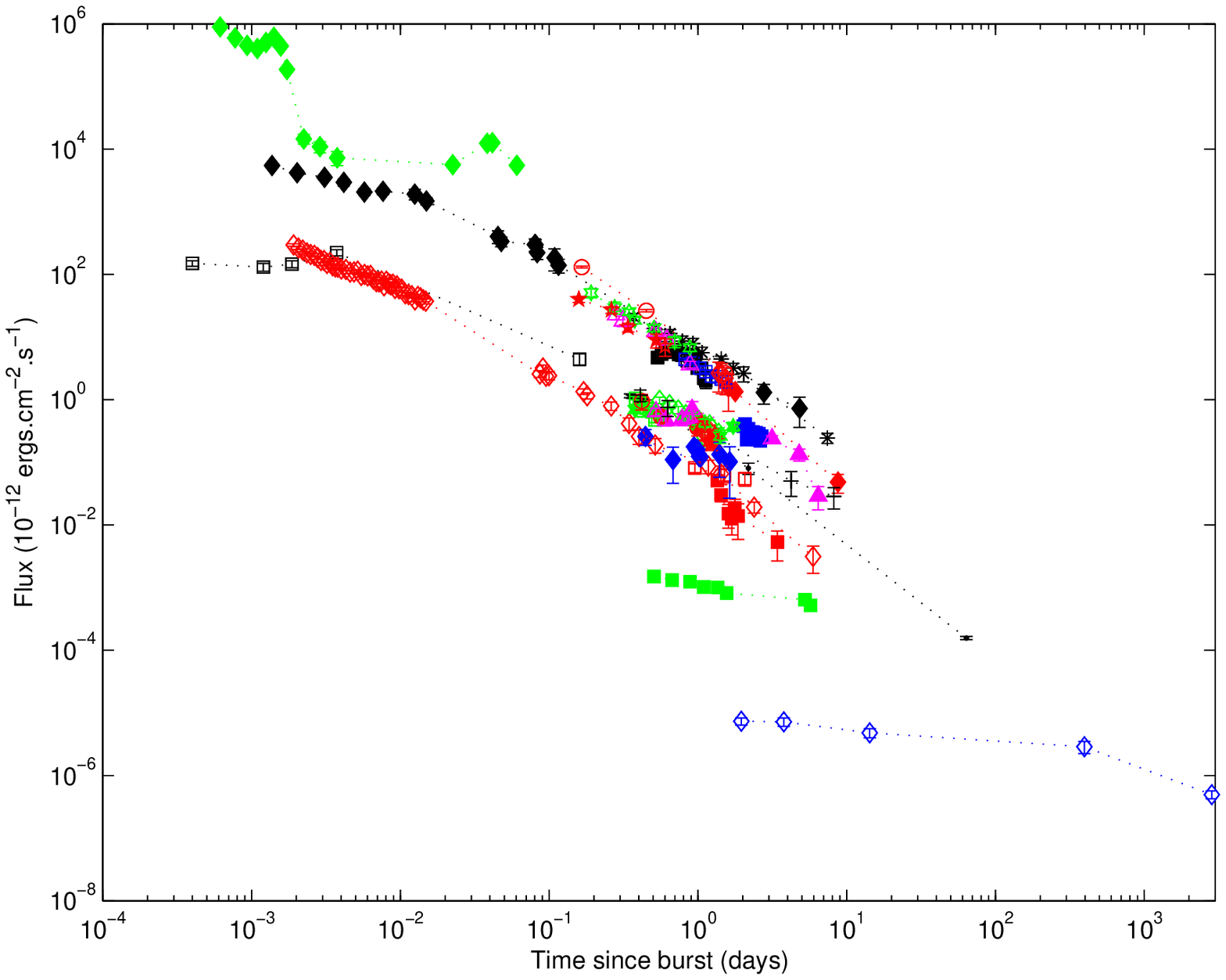}
  \includegraphics[height=.26\textheight]{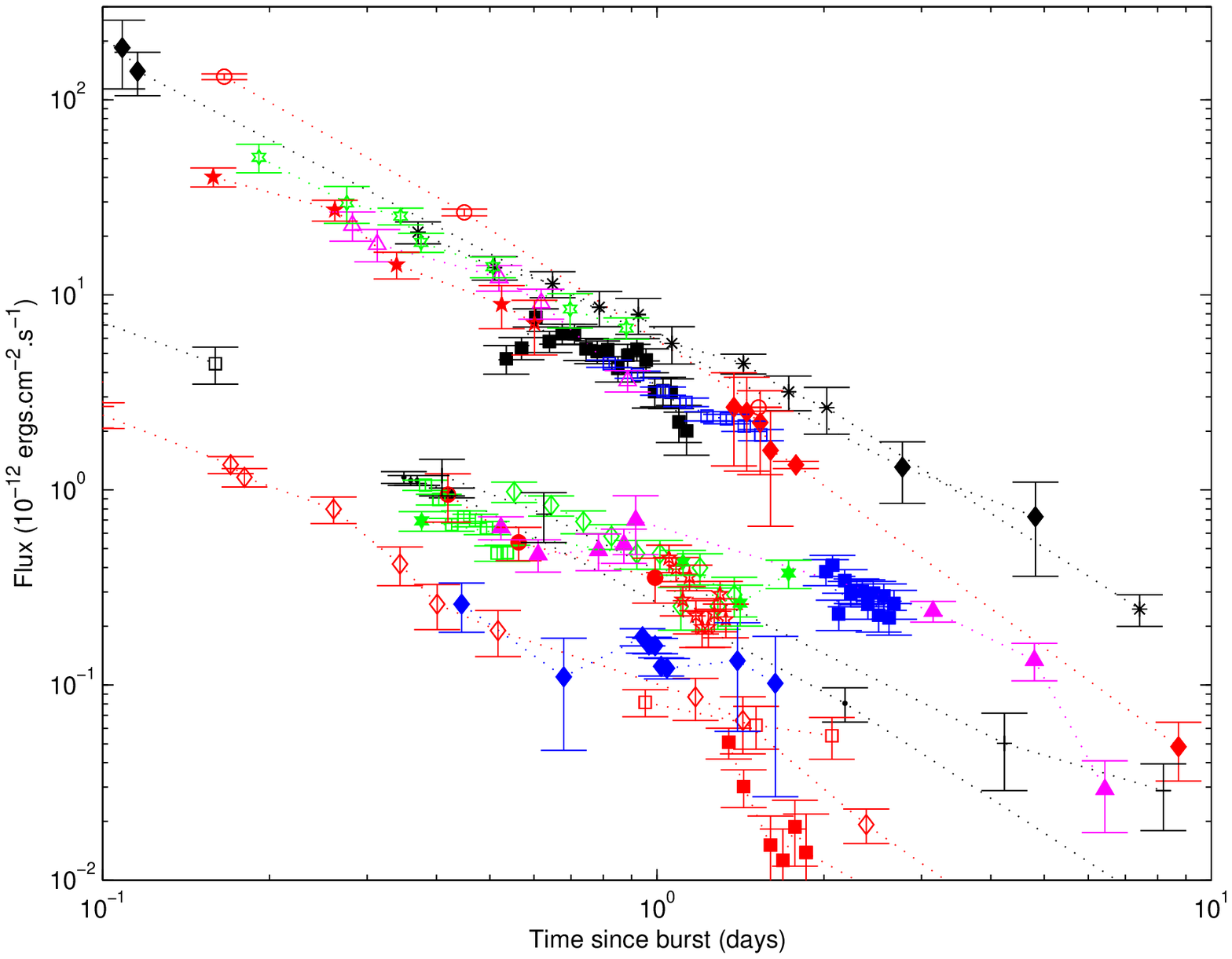}
  \caption{\label{Fig1} The Bo\"er \& Gendre relation made with 26 bursts. Left panel: full scale to show all 
bursts. Right panel: we display the same relation restricted to the time interval from 0.1 to 10 days after 
the GRB.}
\end{figure}

The two groups reported in \cite{boe00} are still present (see Fig. \ref{Fig1}). All but two bursts lie in 
one of the two groups. The only exceptions are GRB 980425 and GRB 031203. In the following we call {\it group 
I} the set of GRB afterglows with the brightest luminosity, and {\it group II} the dimmer ones. The 
probability that a power law luminosity distribution (letting the index be a free parameter) represent the 
observed distribution is, at the maximum, $1.2 \times 10^{-6}$ (index value : -1), thus the observed 
clustering in two groups is very significant.

We computed the mean decay index of the groups. We find $\delta = 1.6 \pm 0.2$ for {\it group I}. If we take 
into account all bursts of {\it group II}, we find $\delta = 1.5 \pm 0.9$. However, if we take into account 
only the bursts with a good decay constraint (hence ignoring GRB 011121 and GRB 030226), we get $\delta = 1.1 
\pm 0.2$. Using a Kolmogorov-Smirnov test to check if this repartition is due to a single population of GRBs, 
we obtain a probability of 0.13 : this distribution of decay indexes may be due to only one population.

\subsection{Validity of the relation as a distance estimator}

Before using this relation as a distance estimator, one may check its validity.

\begin{enumerate}
\item It has been reported by \cite{nar05} a weak clustering in two groups. This effect is probably due to 
the time dilatation correction computed by these authors: As noted above, and also stated by \cite{lia05}, 
this correction needs to be computed without extrapolations (as done in [13]),  otherwise the uncertainties 
on the decay index and on the mean flux level might  broaden the true distribution.
\item Several discrepancies between SWIFT bursts and other bursts have been reported \cite{ber05}. This might 
be the result of a selection effect. While this claim is still valid for the optical afterglows, 
\cite{gen05b} has shown that the X-ray afterglows of Beppo-SAX, XMM-Newton and SWIFT are similar. Thus, no 
selection effect can be objected if one uses X-ray afterglows, as we do.
\item Both nearby and distant bursts may be used to check the validity domain of the relation. At large 
distances (z$>$5) only one burst (GRB 050904) is present in our sample. It presents strong flaring activity 
\cite{wat05, boe06} which may make conclusions difficult to draw at first glance; however this burst  agrees 
with the relation (see Fig. \ref{Fig1}). At small distances (z$<$0.1), two bursts (GRB 980425 and GRB 031203) 
are present in the sample. As can be seen in Fig. \ref{Fig1}, they do not follow the relation. These events 
are the only  outliers we found, and this might be
due to a distance effect : as a conservative hypothesis we prefer to restrict the validity of the Bo\"er \& 
gendre relation to redshifts larger than 0.5.
\end{enumerate}

We emphasize that the bursts were sorted in each groups according only to their X-ray properties (the decay 
index),
and that we did not use any redshift estimate. When the classification was ambiguous (e.g. for a decay index 
of 1.4)we systematically privileged
the group that gave the lowest redshift value, i.e. disfavoring high redshifts (where we want to check the 
relation).

\section{The redshift distribution of pre-SWIFT bursts}

\begin{figure}
  \includegraphics[height=.3\textheight]{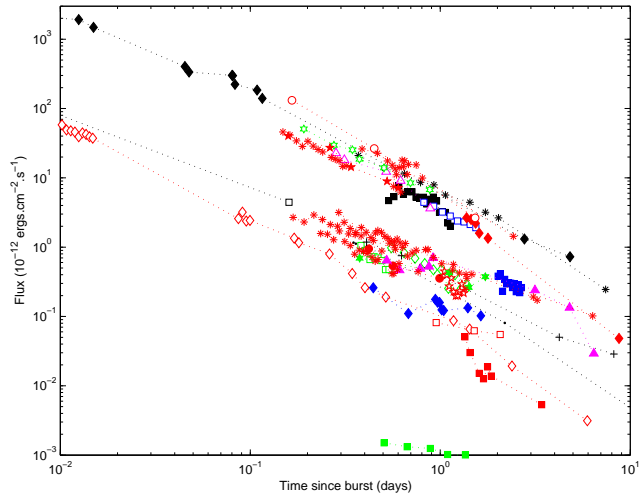}
  \caption{\label{Fig2} The rescaled bursts.}
\end{figure}

\begin{table}
\begin{tabular}{cccc}
\hline
\tablehead{1}{r}{b}{GRB  name}
  & \tablehead{1}{r}{b}{Spectral  index}
  & \tablehead{1}{r}{b}{Observing  satellite}
  & \tablehead{1}{r}{b}{Redshift estimate}   \\
\hline
GRB 001025A& 1.8 & XMM-Newton & 5.8\\
GRB 020322 & 1.1 & XMM-Newton & 1.8\\
GRB 040106 & 0.49 & XMM-Newton & 3.4\\
GRB 040223 & 1.6 & XMM-Newton & 1.9\\
GRB 040827 & 1.3 & XMM-Newton & 1.9\\
GRB 980329 & 1.44 & BeppoSAX & 1.5\\
GRB 980519 & 2.43 & BeppoSAX & 4.8\\
GRB 990704 & 1.68 & BeppoSAX & 1.4\\
GRB 990806 & 1.31 & BeppoSAX & 1.7\\
GRB 001109 & 1.29 & BeppoSAX & 2.8\\
GRB 020410 & 1.3 & BeppoSAX & 0.5\\
\hline
\end{tabular}
\caption{Redshift estimates derived from the Bo\"er \& Gendre relation}
\label{table2}
\end{table}

We used the Bo\"er \& Gendre relation to derive the distance of the bursts listed in Table \ref{table2}. As 
one may note, only one burst does not reach the lower threshold, and thus we do not consider it in the 
following discussion. The mean redshift for SWIFT bursts is 2.7, while the mean pre-SWIFT mean redshift was 
1.2. We took into account the bursts from table \ref{table2} to recomputed the redshift distributions; the 
result is displayed
on the right panel of Fig. \ref{fig0}. As it can be seen, with this method the pre-SWIFT  and  SWIFT 
distributions do agree both at low and high redshifts. The observed difference at intermediate redshift can 
be explained by a lack of SWIFT bursts at these distances, and is compatible with the expectation derived 
from Poisson counting statistics.
We note also the interesting result of GRB 001025A. This burst was observed by XMM-Newton but its optical 
afterglow was never detected, and thus it was classified as dark burst \cite{ped05}. With a calculated 
redshift of $5.8 \pm 0.8$, this can be easily explained by the Lyman alpha cutoff.

As stated above, there is a bias observed with the SWIFT optical afterglows : they appear fainter than the 
pre-SWIFT ones \cite{ber05}. This has strong consequences on the distance estimation of the bursts. This 
estimate is based
on spectroscopic observations of the optical afterglow which is detected mostly after an X-ray observation 
gave the precise
position of the transient. Before SWIFT X-ray observations were made hours after the burst. Hence the optical 
follow-up occurred at least hours, or even days,
 after the event, when the optical transient had significantly faded. Thus the OT detection implied a bright 
source: they are not common as discovered by SWIFT. Because of the cosmological effects, a burst will have 
its flux decreasing with the distance if the afterglow spectral index is larger than 1 (as observed in most 
of the afterglows by [7, 16])  : bright bursts will be on average nearby,  biasing the pre-SWIFT 
distribution against high redshifts. Since our analysis is unbiased in that sense,
we do not see any difference in distance in the two samples.

\section{Conclusions}

We have presented the Bo\"er \& Gendre relation which links the X-ray afterglow luminosity and the GRB source 
distance. We use it to derive the distance of burst sources with unknown redshift. We observe significantly 
more distant bursts by using our method, than selecting them only on their optical properties. We show that 
the redshift distribution of GRB sources computed from the pre-SWIFT and SWIFT sample agree when using the 
Bo\"er and Gendre relation.

We infer that the difference previously reported is due to a selection bias due to the optical measurements 
of the redshift of pre-SWIFT bursts.




\begin{theacknowledgments}
This work was supported by the EU FP5 RTN 'Gamma ray bursts: an enigma and a tool'.
\end{theacknowledgments}






\end{document}